\documentclass[nofootinbib,11pt]{revtex4-1}
\usepackage{amsmath,amssymb,pgf,pgfarrows,pgfnodes,float,appendix, hyperref}
\usepackage{graphicx}
\usepackage{subfigure}
\usepackage[margin=0.9in]{geometry}

\newcommand{\beq}{\begin{equation}}
\newcommand{\eeq}{\end{equation}}
\newcommand{\bea}{\begin{eqnarray}}
\newcommand{\eea}{\end{eqnarray}}

\begin{document}

\title{Conformal Description of Near-Horizon Vacuum States}
\author{Thomas Banks}
\affiliation{Department of Physics and NHETC, Rutgers University, Piscataway, NJ 08854 USA}
\author{Kathryn M. Zurek}
\affiliation{Walter Burke Institute for Theoretical Physics, California Institute of Technology, Pasadena, CA USA }

\begin{abstract}

Motivated by recent work suggesting observably large spacetime fluctuations in the causal development of an empty region of flat space, we conjecture that these metric fluctuations can be quantitatively described in terms of a conformal field theory of near-horizon vacuum states.  One consequence of this conjecture is that fluctuations in the modular Hamiltonian $\Delta K$ of a causal diamond are equal to the entanglement entropy: $\langle \Delta K^2 \rangle = \langle K \rangle = \frac{A(\Sigma_{d-2})}{4 G_d}$, where $A(\Sigma_{d-2})$ is the area of the entangling surface in $d$ dimensions.  Our conjecture applies to flat space, the cosmological horizon of dS, and AdS Ryu-Takayanagi  diamonds, but not to large finite area diamonds in the bulk of AdS. We focus on three pieces of quantitative evidence, from a Randall-Sundrum II braneworld,  from the conformal description of black hole horizons, and from the fluid-gravity correspondence.  Our hypothesis also suggests that a broader range of formal results can be brought to bear on observables in flat and dS spaces.

\end{abstract}

\maketitle
\tableofcontents

\section{Introduction}

Because a detector in any experiment follows a time-like trajectory for a finite proper time, an experimental measurement defines a causal diamond of finite proper time. 
The covariant entropy bound (CEB) asserts that the maximal entropy of any quantum state of the diamond in $d$ dimensions is less than $A(\Sigma_{d-2})/4 G_d$, where $A(\Sigma_{d-2})$ is the area of the bifurcation surface of the diamond.   
Remarkably this bound takes an identical form to the Bekenstein-Hawking entropy, and has been shown to be saturated in several situations where there is no black hole.  For example
\begin{itemize}
\item In any quantum field theory,  tracing out the complement of the region bounded by $\Sigma_{d-2}$ produces a density matrix which is thermal with respect to an operator called the modular Hamiltonian.  In CFTs with an Einstein-Hilbert (EH) dual\footnote{We use the phrase {\it  CFT with an EH dual} to mean a CFT whose correlators can be well approximated by solving (super)gravity equations in AdS space. This is more precise than the conventional {\it large radius holographic dual}.} the thermal and entanglement entropy are $A(\Sigma_{d-2})/4G_d$ \cite{RT,CHM}, where $A(\Sigma_{d-2})$ is the area of a Rindler diamond in AdS space. 

\item In gravity, diamonds in maximally symmetric spacetimes obey a first law of the form $dH_\zeta=-TdS$, where $2\pi T=\kappa$ is the surface gravity of the horizon, $S=A(\Sigma_{d-2})/4G_d$ is the Bekenstein-Hawking entropy, and $H_\zeta$ is the total Hamiltonian generating flow along a conformal Killing vector (CKV) $\zeta$ that preserves the diamond \cite{jv}. The authors of~\cite{jv} argue 
that the modular Hamiltonian for causal diamonds in such spaces is the quantum generator of the CKV that preserves the diamond, evaluated at a point on the diamond's bifurcation surface.  Below, we will embed this generator in a Virasoro algebra. 
\item The entanglement entropy of causal diamonds may in some cases be computed with Euclidean methods. One obtains a finite result $S_{ent}=A(\Sigma_{d-2})/4 G_d$~\cite{Sussug,carlipteitel,Callan:1994py,Cooperman:2013iqr,bdf}.
\item In general  quantum field theories, 
states with finite numbers of local operators acting on the vacuum state exhibit UV divergent entanglement entropy proportional to the area $A(\Sigma_{d-2})$ of the entangling surface~\cite{Srednicki:1993im}. This is interpreted as a renormalization of Newton's constant~\cite{Sussug} if one assumes the Covariant Entropy Bound is saturated for general diamonds.  Note that this point is conceptually distinct from the use of field theory in the AdS/CFT correspondence.  Here, the field theory in question is a bulk effective field theory.
\end{itemize}

\begin{figure}[btp]
\begin{center}
\includegraphics[scale=0.45]{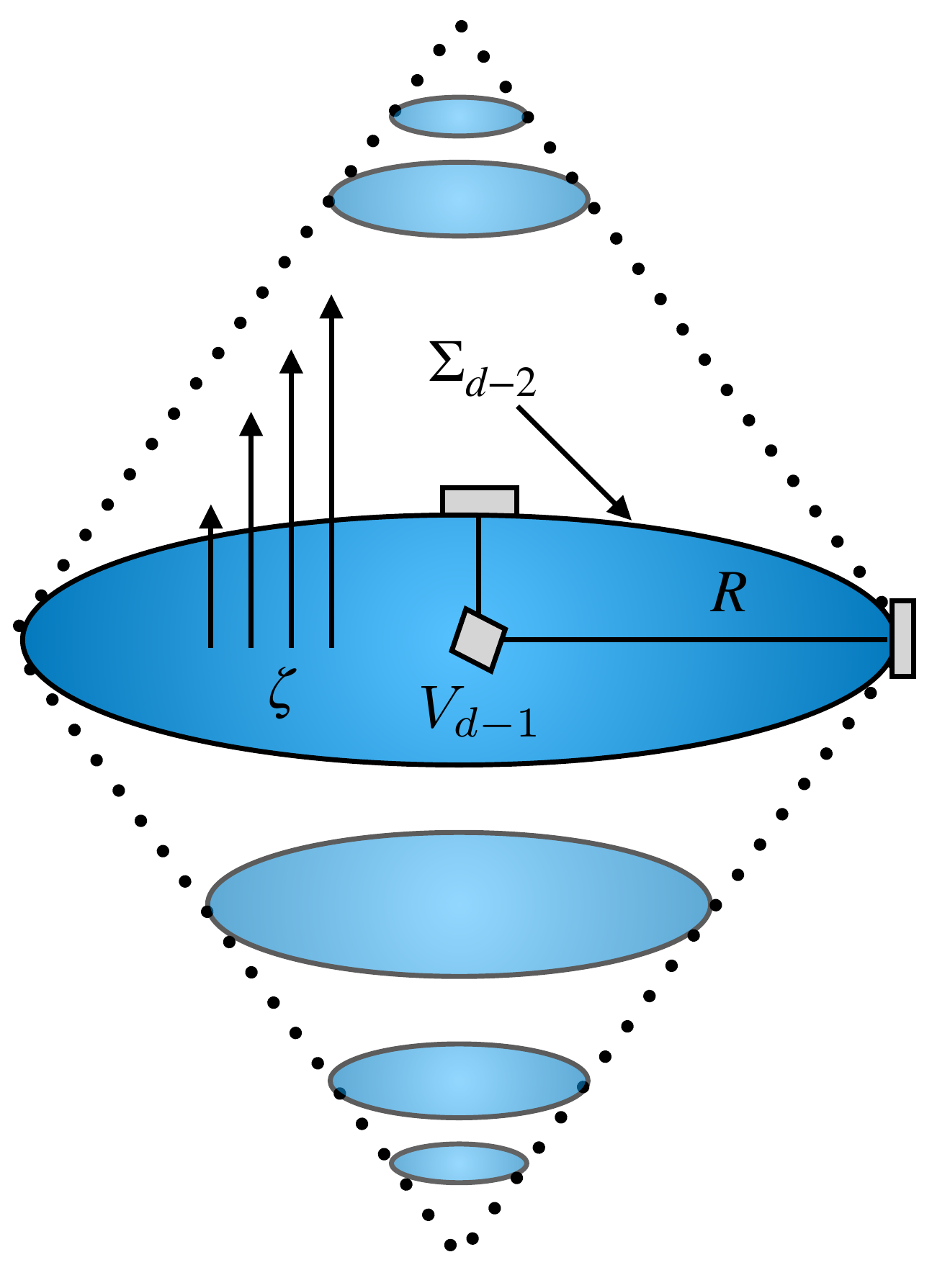}
\label{fig:diamond}
\vspace{-0.5cm}
\caption{Depiction of a causal diamond, with the arms of the interferometer shown on the bifurcate horizon.  The worldlines of the beamsplitter and mirror (not shown for clarity of representation) flow vertically.  The entangling surface is denoted as $\Sigma_{d-2}$, where the far mirrors of the interferometer measure the quantum uncertainty in the position of this surface. The vertical arrows show lines of flow of the conformal Killing vector $\zeta$.}
\end{center}
\end{figure}

These results lead one to consider the idea that the logarithm of the dimension of the Hilbert space of a causal diamond is $\frac{A(\Sigma_{d-2})}{4G_d}$, with $A(\Sigma_{d-2})$ the area of the bifurcation surface defined by light sheets~\cite{bousso1,bousso2,bousso3}.   Let us outline the steps leading to these conclusions in more detail.  The density matrix $\rho$ of any quantum field theory, restricted to such a diamond, is given by (see {\em e.g.} \cite{CHM}) 
\begin{equation}
\rho = \frac{e^{-K}}{\mbox{Tr}(e^{-K})} , 
\label{eq:rho}
\end{equation}
which defines the modular Hamiltonian $K$ of the diamond.  For a CFT, the general form of the modular Hamiltonian (for any state obtained by acting on the vacuum with a finite number of local operators, space-like separated from the diamond) is known in terms of the stress tensor $T_{ab}$ and the CKV $\zeta^b$, $K = H_\zeta = \int dV_{d-1}^a T_{ab} \zeta^b $,  where $V_{d-1}$, shown as shaded disks in Fig.~\ref{fig:diamond}, is a volume element (at fixed time) in the CFT.  Such a density matrix allows one to compute the thermodynamic quantities $\langle K \rangle$ and its fluctuations $\langle \Delta K^2 \rangle$ via the free energy
\beq
F_n = - \frac{1}{n} \log \mbox{ tr}\left(e^{-n K}\right), 
\label{eq:F}
\eeq
such that
\begin{eqnarray}
\langle K \rangle & = & \frac{d}{dn} (n F_n) |_{n=1} \label{eq:K}\\
\langle \Delta K^2 \rangle & = & -\frac{d^2 }{dn^2} (n F_n)|_{n=1} \label{eq:DeltaKsq} .
\end{eqnarray}
Here we have used the ``replica index'' $n$ is a stand-in for the inverse temperature of the system $\beta$.  For continuous $n$ the formula for $F_n$ embeds the modular Hamiltonian in a one parameter set of thermal density matrices. For integer values of $n$ there is a path integral over branched covers of the original manifold, which allows one to perform calculations semi-classically.  The replica trick thus allows one to relate the thermodynamic entropy to geometric variations of the diamond, as considered by many authors, including \cite{Callan:1994py,Lewkowycz:2013nqa,Cooperman:2013iqr}.  In particular, by identifying the Euclidean gravitational effective action by the relation  $n F_n = I_n - n I_1$, one finds the geometric entropy of Ref.~\cite{Callan:1994py} is 
\beq
S_{geom} = \langle K \rangle = \partial_n \left(I_n - n I_1 \right)|_{n = 1},
\eeq
in which case the geometric and entanglement entropies are equivalent.   The geometric entropy has been shown, for the case of an extremal entangling surface, to be ({\em e.g.} \cite{Lewkowycz:2013nqa,Cooperman:2013iqr,xidong})
\beq
S_{geom} = \frac{A(\Sigma_{d-2})}{4 G_d}.
\eeq
Note this quantity should be thought of as a renormalized entropy.  If one imposes a UV cutoff, the infinity is regularized to $\Lambda^{d - 2} A(\Sigma_{d-2})$.  It was proposed in~\cite{Sussug} that this infinity could be absorbed in a universal way into the definition of Newton's constant $G_d$, in any theory of quantum gravity well-approximated by the Einstein-Hilbert action.  This formula originated in the work of Carlip and Teitelboim \cite{carlipteitel} and of Susskind \cite{Sussug}, and was also anticipated in earlier work of Srednicki \cite{Srednicki:1993im}. 
The same relation holds in AdS/CFT for the entanglement entropy of a causal diamond anchored to the boundary \cite{Ryu:2006bv}
\beq
\langle K \rangle = S_{ent} = \frac{A(\Sigma_{d-2})}{4 G},
\label{eq:RT}
\eeq
where $A(\Sigma_{d-2})$ is now the area of the bifurcate horizon in the bulk (known as the Ryu-Takayanagi (RT) surface).

Since the number of degrees  of freedom at {\em large} scales is smaller than what would be naively expected from Quantum Field Theory, this naturally suggests that there may be correlations between the degrees-of-freedom.   It has been suggested that such correlations could be observable in modern interferometers, mostly based on heuristic holographic arguments and matrix theory \cite{Hogan1,Hogan2,Chou:2017zpk}.   The first realistic theoretical proposal for calculating fluctuations of observables in finite causal diamonds (such as interferometers measure), utilizing modern holographic techniques, was made in Refs.~\cite{VZ1,VZ2}, and an effective field theory description was formulated in Ref.~\cite{pixellon}.  It is easy to understand the relevance of a finite size causal diamond to an interferometer experiment -- a single pass of the interferometer defines a causal diamond in space-time, with the geodesic of the beam splitter following the center of the diamond from tip-to-tip and the far mirror intersecting the diamond at the outside corner.  This is shown in Fig.~\ref{fig:diamond} (see also the figure in Ref.~\cite{VZ1}).

The results of \cite{VZ2} were particularly interesting, because they revealed an unanticipated universality in the ratio between the fluctuations of entanglement entropy\footnote{In the literature, the fluctuations of the modular Hamiltonian have been called {\it the capacity of entanglement}~\cite{deBoer:2018mzv,Nakaguchi:2016zqi}.  We will use this term, and the phrase {\it modular fluctuations} interchangeably.} and the entanglement entropy itself. 
For any system with an entanglement spectrum\footnote{The entanglement spectrum is the eigenvalue spectrum of the modular Hamiltonian.} whose Laplace transform is dominated by a Gaussian peak\footnote{An example where there is no such Gaussian peak is a finite dimensional system, whose large eigenvalue spectrum grows like $n(E) \sim e^{E^p}$ with $p > 1$. The stationary point of the Laplace transform integrand is a minimum, rather than a maximum of the eigenvalue distribution.}, we expect a relation
\begin{equation} 
\langle \Delta K \rangle^2 = \alpha \langle K \rangle.
\label{eq:alpha}
\end{equation}  
The authors of~\cite{VZ2} found that, for RT diamonds, the bulk dual of a causal diamond on the boundary in a conformal field theory with an Einstein-Hilbert (EH) dual, 
\beq
\alpha = 1.
\eeq
This is independent of all details of the CFT, including its dimension.   Although any ${\cal O}(1)$ value of $\alpha$ is interesting, the observation of a universal value would be both phenomenologically and theoretically striking.  It provides a precise target for experimental investigation, and also gives important information about the general principles underlying a theory of quantum gravity.

The result of \cite{VZ2} was obtained by using black hole entropy formulae for topological black holes, obtained by small (replica) deformations of AdS/Rindler space.  It is notable that the result is confirmed by an independent calculation.  Perlmutter~\cite{perl}, using the observations of~\cite{CHM,HMSY} that Renyi entropies of spherical diamonds in CFTs could be computed as thermal entropies of the same CFT in hyperbolic space, showed that for a general CFT $\alpha $ was proportional to the coefficient in the stress tensor two-point function.  When applied to CFTs with an Einstein-Hilbert dual, this formula gives $\alpha = 1$~\cite{perlpriv}. 

Because no experiment exists in the bulk of AdS, we need to be able to calculate modular fluctuations in causal diamonds small compared to the radius defined by the cosmological constant (c.c.).   We will carry out a few simple calculations, and make a few observations, to make the case for a universal value of $\alpha = 1$ in small ({\em i.e.} Minkowski) and RT diamonds.  We will argue that we expect the same result to hold for the cosmological horizon in dS.
\begin{enumerate}
\item First, we calculate $\langle \Delta K^2 \rangle$ on a flat Randall-Sundrum brane~\cite{RSII}.  The Randall-Sundrum model relates gravity weakly coupled to a CFT with an EH dual, to the physics on an infinite flat brane in AdS space.  This simple model for QG in flat space gives $\alpha = 1$.
\item Second, we make a conjecture stemming from seminal work of Carlip and Solodukhin~\cite{Carlip:1994gy,Carlip:1998wz,Carlip:1998qw,carlipreview,solo} relating the entropy of black holes to the representation theory of a near-horizon Virasoro algebra. We will generalize this conjecture to small empty diamonds and the cosmological horizon of dS space.  It states that the modular Hamiltonian of a diamond has an entanglement spectrum identical to that of the asymptotic $L_0$ generator in a $1 + 1$ dimensional CFT where the Cardy formula is correct.   In fact, the result $\alpha = 1$ depends only on the scaling of the log density of states with $L_0$ eigenvalue, 
\begin{equation}
{\rm ln} \rho(L_0) = B \sqrt{L_0},
\end{equation}  
with $B$ a large constant.  Below we will call the eigenvalues of the $L_0$ generator $E I$, where $E$ is the energy of the CFT, quantized on an interval of length $I$.
The full Cardy formula is necessary to get the exact result for the entanglement entropy~\cite{carlipreview,solo}. 
\item Lastly, the fluid gravity correspondence: a large number of works over the years~\cite{Damour:1978cg,Damour:1979wya,bredberg1,bredberg2,liu,rangamani,Compere:2011dx} (and references therein) have shown that near-horizon Einstein equations, subjected to a few global boundary conditions, reduce to the Navier-Stokes (NS) equation, with universal coefficients.  In particular~\cite{Bredberg:2010ky} showed that the ratio of shear viscosity to entropy density for small diamonds, is universal and agreed with that found in~\cite{Policastro:2002se,Policastro:2002tn} for large black holes in AdS. The fluctuation dissipation theorem (FDT) then tells us that fluctuations should be universal.  The fluctuations relevant to the incompressible NS equation are
those of the $0i$ components of the Brown-York stress tensor, evaluated on the ``stretched horizon," a time-like ``hyperboloid" that approaches the bifurcation surface of the horizon at a minimal distance of one Planck unit.  These entropy fluctuations are {\em not} the main focus of the present paper, but their universality gives credence to the idea that entropy fluctuations are also universal.  
\end{enumerate}

One way of understanding our results is as a conformal description of near-horizon states.  The near-horizon limit will be taken for the causal diamond described by a metric with ``blackening factor'' $f(r)$,
\begin{equation}
ds^2 = -  f(r) dt^2 +  \frac{r^2}{f(r)}+ r^2 d\Sigma_{d-2}^2 .
\label{eq:topologicalBH}
\end{equation}
This metric covers a causal diamond with different blackening factors depending on the background.  For example, 
in empty Minkowski, 
$f(r) = 1-r/R$ \cite{VZ1}, where $R$ is the radius of the bifurcation surface of the diamond boundary, and for boundary anchored diamonds in AdS, $f(r) = r^2/L^2-1$, where $L$ is the AdS curvature.

In the near-horizon limit, we can incorporate metric fluctuations by linearizing the blackening factor about the entangling surface at some position $r_n$: 
\begin{equation} 
ds^2 = - 4\pi T_n(r - r_n) dt^2 +  \frac{dr^2}{ 4\pi T_n (r - r_n)}  + r_n^2 d\Sigma_{d-2}^2 , 
\label{eq:linearmetric} 
\end{equation} 
where the temperature is $T_n = f'(r_n)/4\pi$.   In the language of~\cite{xidong}, $r_n$ is the position of a cosmic $d - 2$ brane, which sources the fluctuation of the geometry.  This mirrors the argument of~\cite{VZ1} which identifies the fluctuations as due to a gravitating mass distribution near the horizon of the classical geometry.  For the case of boundary anchored diamonds in AdS, the dependence of $r_n$ on $T_n$ can be computed by insisting that when $T_1 = n T_n$ the global geometry be asymptotic to an $n-$fold cover of the boundary, branched over the sphere to which the RT surface asymptotes.   Throughout this paper we will refer to the undeformed geometries having $n=1$ to have a horizon radius $r_{n=1} \equiv r_0$ and with corresponding horizon temperature $T_0 = \beta_0^{-1}$.

In the near-horizon limit, we will see, following the analysis of Solodukhin and as discussed in point 2 of the enumerated list above, that the $d$-dimensional Einstein-Hilbert action can be reduced to a classical 2-d conformal field theory on the light cone directions.  A plausible assumption about how the algebra of that CFT is realized in the quantum theory will allow us to obtain $\alpha = 1$.  We will argue that such a calculation will accurately reproduce the global near-horizon behavior when the proper time for an inertial diamond observer is of the order of, or larger than, the radius $r_n$ of the entangling surface.  We will see that this criterion is met for RT and small diamonds in AdS and the cosmological horizon of dS, but not for large diamonds in AdS.  

More importantly, this assumption also offers concrete paths forward for modeling the near-horizon behavior of quantum gravity even in ordinary flat space, utilizing the beautiful and powerful techniques of 2-d CFTs.  We reserve further work along these directions for the future.  The outline of this paper follows the numbered list above.  In the next section we show how $\langle \Delta K^2 \rangle = \langle K \rangle = \frac{A({\cal B})}{4 G_d}$ in a spherical ball ${\cal B}$ on a flat RS-II brane having Newton's constant $G_d$.  In Sec.~\ref{sec:SecIII}, we consider that this same result is derived from positing that the near-horizon dynamics of a causal diamond is described by a 2-d CFT with density of states given by the Cardy formula. We also relate our conjecture of universal entropy fluctuations to~\cite{Bredberg:2010ky}, which showed that the famous result~\cite{kovtun} that the viscosity to entropy density ratio for large black holes in AdS is $\frac{\eta}{s} = \frac{1}{4\pi}$, valid for any weakly curved null surface in a model satisfying the Einstein equations.    In Sec.~\ref{sec:largediamonds} we discuss why the results of Secs.~\ref{sec:braneworld}-\ref{sec:SecIII} are not expected to apply to large diamonds in the bulk of AdS.  Finally, in Sec.~\ref{sec:nested}, we show how entropy fluctuations give rise to fluctuations in the time-of-arrival in a light beam that traces out the causal development of ${\cal B}$, by following a series of nested causal diamonds.  Though utilizing distinct methods, these results on length fluctuations exactly agree with Ref.~\cite{VZ2}.

\section{RS-II Flat Brane in AdS/CFT}
\label{sec:braneworld}

\begin{figure}[btp]
\begin{center}
\includegraphics[scale=0.6]{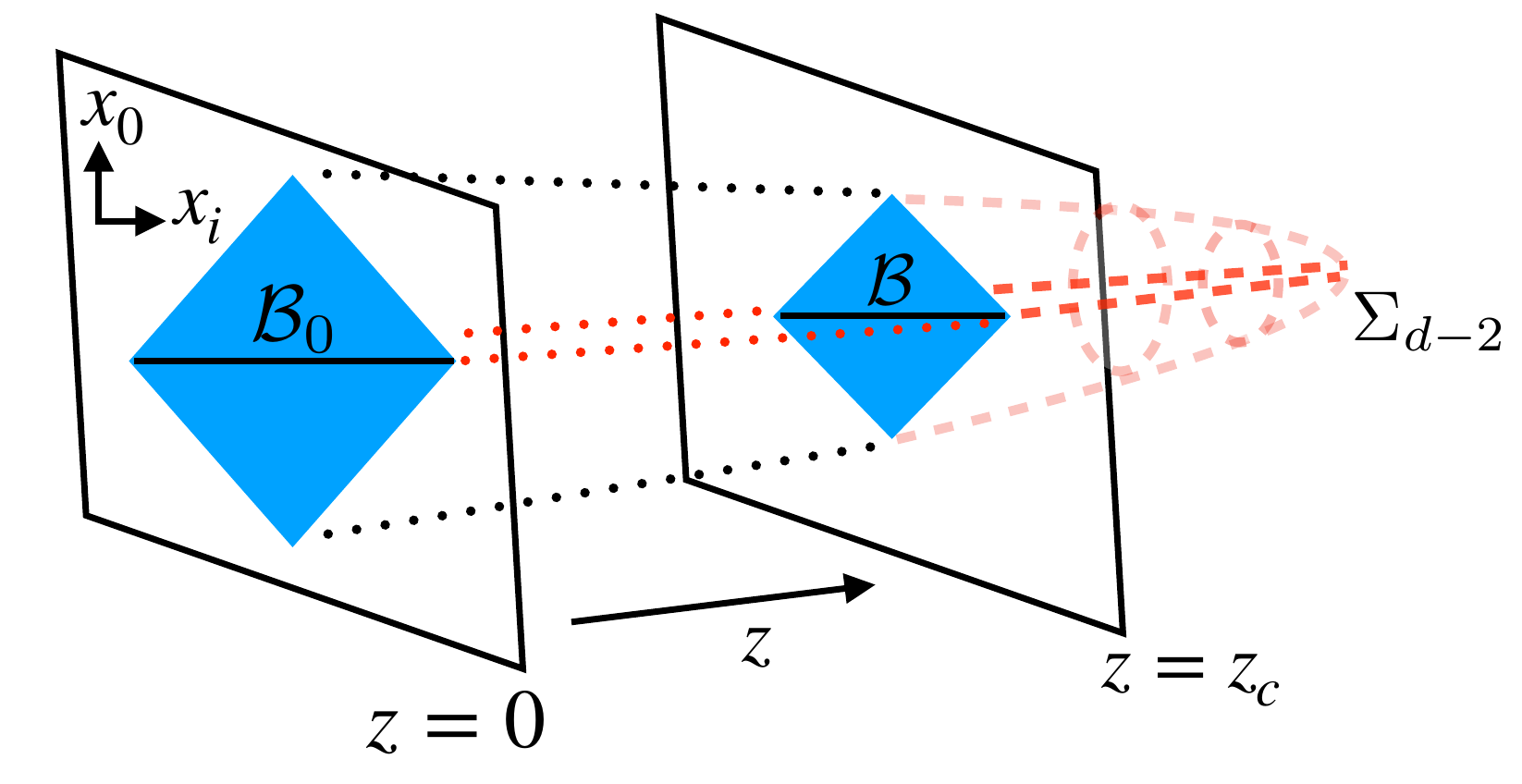}
\label{fig:RSII}
\vspace{-0.5cm}
\caption{Depiction of the RS-II braneworld scenario.  The boundary is located at $z = 0$ and the RS-II brane at $z = z_c$.  We compute the modular fluctuations $\langle \Delta K^2 \rangle$ in the volume enclosed by ${\cal B}$ at the bifurcate horizon on the brane causal diamond, utilizing the holographic map from the bulk RT surface, shown in darker red and labeled $\Sigma_{d-2}$.}
\end{center}
\end{figure}

We begin by considering a braneworld set-up as shown in Fig.~\ref{fig:RSII}.  For continuity, our notation largely follows that of Ref.~\cite{VZ2}.  Our goal is to compute the modular energy $\langle K \rangle$ and its fluctuations $\langle \Delta K^2 \rangle$ in the volume enclosed by the spherical entangling surface ${\cal B}$ on the brane, utilizing a bulk holographic calculation. We begin with  a $d=D+1$ dimensional bulk AdS space with AdS curvature $L$ and metric in Poincare coordinates
\begin{equation}
ds^2 = L^2 \frac{dz^2 + dx_i^2 - dx_0^2}{z^2}
\end{equation}
with $i = 1,....,D-1$.  We take the flat $D$-dimensional brane located at some position $z_c$ in the bulk, as also shown in the figure.   In such a set-up, gravity is induced on the brane, with a dimensionally reduced Newton constant ({\em e.g.} \cite{Myers:2013lva})
\begin{equation}
G_D = \frac{(D-2) }{2 L}G_d.
\label{eq:G}
\end{equation}

The measuring apparatus is confined to the brane (labeled by ${\cal B}$), with light beams from a beam splitter (shown at the bottom of the brane causal diamond) reflecting off a mirror at the bifurcate horizon on the brane, and recombining again at the top of the causal diamond.  
The causal diamond on the brane anchors the Ryu-Takayanagi (RT) surface $\Sigma_{d-2}$ in the bulk.  This spatial slice of $V_{d-1}$ (shown as shaded) is given by $z^2 + x_i^2 \leq R^2$, and the full interior of the bulk diamond is given by
\begin{equation}
R^2 - z^2 - x_i^2 + x_0^2 \geq 2 R |x_0|.
\end{equation}
The modular fluctuations $\langle \Delta K^2 \rangle$ can also be calculated in the bulk \cite{VZ2}, with the result $\alpha = 1$, as noted in the introduction. In the bulk of AdS, Ref.~\cite{VZ2} showed that these modular fluctuations gravitate and induce metric fluctuations.  

Because gravity is induced on the brane, with strength given by Eq.~\ref{eq:G}, an instrument on a brane at $z_c$ will also experience modular energy fluctuations.  To relate the bulk and brane modular fluctuations, we must geometrically relate the transverse area of the surface on the brane $A({\cal B})$ to the RT bulk surface $\Sigma_{d-2}$ which is anchored to it.  

To do this, it is useful to put the metric on the interior of the bulk causal diamond into AdS-Rindler form, Eq.~\ref{eq:topologicalBH}.
Here the transverse plane (having the entangling surface) can be written as
\begin{equation}
d\Sigma_{d-2}^2 = d \rho^2 + \sinh^2 \rho d\Omega_{D-2}^2.
\end{equation}
The bulk light cone directions $r,~t$ are related to the Poincare coordinates via
\begin{equation}
\frac{R^2 - z^2 - x_i^2 + x_0^2}{2 R z} = \left(\frac{r^2}{L^2} - 1\right)^{1/2} \cosh \frac{t}{L},~~~~~~~\frac{x_0}{z} = \left(\frac{r^2}{L^2} - 1\right)^{1/2}\sinh \frac{t}{L},
\end{equation}
while the transverse direction is
\begin{equation}
|x_i| = R \tanh \rho.
\label{eq:xi}
\end{equation}
The blackening function is of the standard (topological) black hole form
\beq
f(r) = \left(\frac{r^2}{L^2} - 1\right) + \frac{4 \pi M}{(d-2)}\frac{L^{d-2}}{r^{d-3}} \frac{4 G_d}{A(\Sigma_{d-2})},
\label{eq:blackening}
\eeq
where the black hole mass is replaced with the modular fluctuations $M = \frac{\Delta K}{2\pi L}$, following Ref.~\cite{VZ2}.  These modular fluctuations hence cause geometric fluctuations.  We seek to understand the relation between the modular fluctuations computed in the bulk and those induced on the brane.  

To do this, we relate the area appearing in Eq.~\ref{eq:RT} to the area of the entangling surface on the brane $A({\cal B})$.
Now the area of the RT surface is
\begin{equation}
A(\Sigma_{d-2}) = L^{D-1} \Omega_{D-2} \int_0^{\rho_c} d\rho \sinh^{D-2}\rho.
\label{eq:RTarea}
\end{equation}
The location of the entangling surface is found by noting that the transformations between Poincare and topological BH coordinates imply that the bulk diamond satisfies
\begin{equation}
\left(\frac{(R-x_0)^2 - z^2 - x_i^2}{2 R z}\right)\left(\frac{(R+x_0)^2 - z^2 - x_i^2}{2 R z}\right) = \frac{r^2}{L^2} - 1.
\end{equation}
We can thus see that $\Sigma_{d-2}$ is located at $x_0 = 0$ and $z^2 + x_i^2 = R^2$.   We also find that fixing $z_c $ is equivalent to fixing the brane location in the bulk according to the relation
\begin{equation}
z_c = R\sqrt{1 - \tanh^2 \rho_c} = \frac{R}{\cosh \rho_c},
\end{equation}
such that $z_c \ll R$ is equivalent to $\rho_c \gg 1$.

Since the bulk RT diamond is anchored to the boundary diamond, a light beam traveling from tip-to-tip on the boundary diamond will arrive at the same time as a light beam traveling from tip-to-tip on the bulk RT diamond.  This time is given by 
\begin{equation}
T =\frac{L}{z_c}(R-z_c) =  L(\cosh \rho_c - 1) \approx \frac{e^{\rho_c}}{2} L.
\end{equation}

 Then, combining Eq.~\ref{eq:RTarea} with Eq.~\ref{eq:G}, we have
\begin{equation}
\frac{A(\Sigma_{d-2})}{4 G_d} \simeq \frac{L^{D-2} \Omega_{D-2}}{4 G_{D}} \left(\frac{e^{\rho_c}}{2}\right)^{D-2},
\end{equation}
where we are taking the $\rho_c \gg 1$ limit.   We now identify 
\beq
A({\cal B}) = L^{D-2} \Omega_{D-2} \left(\frac{e^{\rho_c}}{2}\right)^{D-2},
\eeq 
as the area of $D-2$ dimensional surface ${\cal B}$.
Therefore, utilizing the $\alpha = 1$ result for bulk RT diamonds of Ref.~\cite{VZ2}, we have 
\begin{equation}
\langle \Delta K^2 \rangle = \frac{A(\Sigma_{d-2})}{4 G_d} = \frac{A({\cal B})}{4 G_D}.
\end{equation}
This suggests that modular Hamiltonian fluctuations in Minkowski space are the same as in AdS, {\em i.e.} $\alpha = 1$.  

Note that very similar reasoning, applied now to $\langle K \rangle$ rather than $\langle \Delta K^2 \rangle$, gives
\beq
\langle K \rangle = \frac{A(\Sigma_{d-2})}{4 G_d} = \frac{A({\cal B})}{4 G_D}.
\eeq
The vacuum entanglement entropy of a causal diamond on an RS-II brane was also calculated by independent methods in Ref.~\cite{Myers:2013lva}, and the same result was obtained, supporting the results obtained here.

At first sight $\langle \Delta K^2 \rangle = \langle K \rangle$ for Minkowski diamonds may seem surprising, so we explore next why this result could have been expected from the dynamics of conformal field theories in the near-horizon limit.

\section{Near-Horizon Dynamics}
\label{sec:SecIII}

We first consider a general coordinate system that covers a causal diamond characterized by a geodesic proper time $T_{prop}$.  There is a different near-horizon behavior for large finite diamonds than for boundary-anchored and small diamonds in AdS, or the cosmological horizon of dS space.   The essential difference is simple, and determined by whether the inequality
\beq
T_{prop}^{d-2} \gtrsim A(\Sigma_{d-2})
\label{eq:Tprop}
\eeq
is satisfied.  If it is, our claim is that there is sufficient time for the near-horizon dynamics to equilibrate the state of the system, and $\alpha = 1$.
\begin{enumerate}
\item For diamonds smaller than the curvature radius $L$, area scales like $T_{prop}^{d-2}$ and this scaling relation is valid for the proper time along any trajectory of finite acceleration.  
\item For the cosmological horizon of dS the proper time along any timelike trajectory connecting the tips of the causal patch can be arbitrarily long, while the area is bounded by $L^{d-2}$.  Thus, localized probes of the horizon measure the equilibrium behavior of the entire diamond.  
\item For boundary-anchored diamonds in AdS, the same is true, though in this case the system has infinite entropy.   
\end{enumerate}
In all three of these cases, Eq.~\ref{eq:Tprop} is satisfied.  By contrast, for finite area diamonds in AdS, the proper time along any trajectory is bounded by $\frac{L\pi}{2}$, while the area can be arbitrarily large.  Although the local near-horizon dynamics is identical to that of the other kinds of diamonds, no diamond observer can follow the system long enough to probe the equilibrium state. The calculations of this section are valid locally on all horizons with the near-horizon Rindler factorization exhibited below.

The authors of~\cite{hp,ss} have argued that the behavior of the quasi-normal modes on Minkowksi diamonds indicates that the quantum systems responsible for the equilibration are in fact fast scramblers. The same arguments apply to the cosmological horizon of dS space, and indeed to the local behavior on any diamond horizon.  Fast scrambling systems have a characteristic time scale $t_0$ and become completely equilibrated in a time $t_0 {\rm ln} S$ where $S$ is the total number of q-bits in the system.  For small black holes $t_0$ is the Schwarzschild radius, while for the dS horizon it is the Hubble time.  For AdS-Rindler $t_0$ is the AdS radius $L$.   In all three of the cases enumerated above, $t_0 {\rm ln}\ S$ is of the same order or smaller than the proper time along trajectories in the diamond and one can explore the equilibration of perturbations over the entire horizon within the proper time available to a detector on such a trajectory. For AdS-Rindler we must cut off the $z$ coordinate on the hyperboloid, similarly to the RS-II model, to render the entropy finite in order to make this statement unambiguous.

In the case of large AdS diamonds/black holes, however, the local quasi-normal modes dissipate after having spread to only a distance $\sim L$ on the horizon, which is much smaller than their Schwarzschild radius. Furthermore, as we will recall below, the quasi-normal spectrum of large AdS diamonds includes sound modes with wavelengths $> L$, which indicate ballistic propagation of energy and entropy densities, rather than fast scrambling.  By contrast, the quasi-normal spectrum of all other horizons, including AdS-Rindler space does not have sound modes.  We believe that this is an indication that our local calculation of the modular fluctuations is valid for any causal diamond with a Rindler horizon, but that the thermal equilibrium ensemble dual to a large radius AdS causal diamond can have a different value of $\alpha$ because its entropy is dominated by modes that propagate ballistically on the large horizon.  This is of course the message of the AdS/CFT correspondence.  The equilibrium dynamics is that of a local field theory on the sphere.

Our claim, following the work of Carlip~\cite{carlipreview} and Solodukhin~\cite{solo}, is that locally on the bifurcation surface of a diamond with a Rindler near-horizon expansion, the quantum dynamics is governed by a Virasoro algebra in a representation satisfying Cardy's theorem.  
The argument begins with a rigorous classical theorem that the algebra of near-horizon symmetries consists of conformal transformations of the flat two dimensional Rindler space,\footnote{The authors of Ref.~\cite{carlipreview,solo} emphasize that there are two such Virasoro algebras for a time symmetric situation, which are time reflections of each other. Only one is apparent at a given point on the horizon. The CFT on the past/future portion of the diamond boundary has only incoming/outgoing modes.} which acts on the horizons of all of these diamonds in space-time.  It is important to note that the $L_0$ generator of this algebra is precisely the CKV of the background geometry which switches from time-like to space-like on the diamond boundary.  The quantum generator associated with the action of this CKV on the bifurcation surface of the diamond was proposed~\cite{CHM,jv} to be the modular Hamiltonian of the diamond.  It has been shown \cite{carlipreview,solo} that, using the symplectic structure derived from the EH Lagrangian, the Poisson brackets of these generators have a central charge and there is a classical value for the $L_0$ generator, coming from the boundary terms.   If one {\it assumes} that the quantum model of horizons is a representation of the Virasoro algebra satisfying the constraints giving rise to the Cardy formula for the asymptotic density of $L_0$ eigenvalues, then that formula reproduces the Bekenstein-Hawking area law for black holes\footnote{In fact, the argument works for rotating black holes as well as the spherically symmetric geometries considered here.}.

We will summarize and generalize these arguments using the formalism of Solodukhin~\cite{solo}, who dimensionally reduced the Einstein-Hilbert action on a class of metrics of the form
\begin{equation} 
ds^2 = g_{ab} dy^a dy^b + \rho^2 (y) d\Sigma_{d-2}^2  = \rho^2 (y) [\rho^{-2} g_{ab} dy^a dy^b + d\Sigma_{d-2}^2 ]. 
\end{equation}
Solodukhin assumed that $d\Sigma_{d-2}^2$ was the round metric on a sphere.
We will argue that in fact  $d\Sigma_{d-2}^2 $ can be any fixed $d-2$ dimensional Riemannian manifold, and $g_{ab}$ a two dimensional Lorentzian metric, which we will soon restrict to be Rindler space.  Evaluated on such a configuration, the EH action reduces to a metric for dilaton gravity, with the dilaton a function of $\rho$.  Such metrics, in the near-horizon limit, can be expanded, following Solodukhin, in the form of Eq.~\ref{eq:linearmetric}.  For our purposes, the transverse metric $d\Sigma_{d-2}^2$ can be any Riemannian manifold, for which this reduction produces a valid solution of the $d$ dimensional Einstein equations.
To see the reduction, note that the metric $\tilde{g}$ in square brackets is a product manifold so that
\begin{equation} 
R_d (\tilde{g}_{\mu\nu} ) = R_2 (\rho^{-2} g_{ab} ) + R_{d - 2} . 
\end{equation}
Our metric is conformal to this one so that~\cite{carroll} 
\begin{equation} 
R(g_{\mu\nu}) = \rho^{-2} R_d(\tilde{g}_{\mu\nu} ) + 2\rho^{-1} (d - 1) \tilde{g}^{ab} \tilde{\nabla}_a \tilde{\nabla}_b (\rho^{-1}) - d(d - 1)  \tilde{g}^{ab} \partial_a \rho^{-1} \partial_b \rho^{-1} . 
\end{equation}  
Here we have used the fact that the conformal factor depends only on the two dimensional coordinates, to simplify the general relation between scalar curvatures of conformally related metrics.  

Combining the last two equations, we see that the geometry of the $d - 2$ dimensional bifurcation surface enters into the two dimensional Lagrangian only via overall factors of the area of the surface and its integrated scalar curvature.  Furthermore, the $d - 2$ dimensional curvature term multiplies something that is a pure potential for the dilaton, with no derivatives of the two dimensional fields.    As we will see below, Solodukhin argues that such potential terms become irrelevant near the horizon, where the theory thus becomes conformal.  

   {\it The upshot of these remarks is that we can translate Solodukhin's analysis of spherically symmetric black hole horizons to the bifurcate horizon of a causal diamond in a Lorentzian space-time solving the Einstein equations.}  Moreover, the argument that the curvature of the bifurcation surface is negligible near the horizon implies that the calculation is ultra-local on the horizon.  This is quite analogous to the replica calculation of $\langle K \rangle$ and $\langle \Delta K^2 \rangle$.  

 We now recapitulate Solodukhin's analysis of the spherically symmetric case, where he has carried out the algebra of the general discussion of the previous paragraphs. Solodukhin writes the dimensionally reduced action as\footnote{This is Eqn. 5.3 of~\cite{solo}, with the substitution $r\rightarrow \rho$ since we use $r$ to denote a coordinate rather than a field.} 
\begin{equation}
I = -\frac{\Sigma_{d-2}}{16\pi G_d} \int d^2 y\ \sqrt{- g_2} [\rho^{d - 2} R_2 (g_{ab})+ (d - 3) (d -2) \rho^{d - 4} (\nabla \rho)^2  + (d - 3)(d - 2) \rho^{d-4} ]  . 
\end{equation} 
Note that a $d$ dimensional cosmological constant would similarly add a potential term that is a power of $\rho$ .  The field $\rho$ is the radius of the $\Sigma^{d-2}$ entangling surface. 
When we solve its equation of motion it will have the value $\rho_0$ at the horizon.    

Now let us suppose that there is a solution of the $d$ dimensional equations, which has a causal diamond whose bifurcation surface has fixed area $\Sigma_{d-2}$, in Planck units.  Near the bifurcation surface, generically, the two-dimensional metric will be two dimensional Rindler space as in Eq.~\ref{eq:linearmetric}.
Following Solodukhin, define 
\begin{equation} 
\rho^{d-2} =  \rho_0^{d-2} \sqrt{\frac{2 \pi  }{ S}\frac{(d - 3)}{ (d - 2)}}q \phi .
\label{eq:rhophi}
\end{equation} 
where $S = \Sigma_{d-2} \rho_0^{d-2}/4 G_d$ and $q$ is an arbitrary parameter.  We immediately see that the dilaton $\phi$ is related to the area.

One then redefines the background metric by a $\phi$ dependent Weyl transformation
\begin{equation} 
g_{ab} = (\frac{\phi_0}{\phi})^{\frac{d-3}{d-2}} e^{\frac{2\phi}{ \Phi_0}} \bar{g}_{ab}, 
\end{equation}   
where $\Phi_0 = \sqrt{\frac{2 S}{\pi}\frac{(d-3)}{(d-2)}} = 2 q (d-3)/(d-2) \phi_0$, with $\phi_0$ is the value of $\phi$ near the horizon according to the definition in Eq.~\ref{eq:rhophi}. The action takes the form 
\begin{equation} 
I = - \int d^2 y \sqrt{-\bar{g_2}} \left[\frac{(\nabla\phi)^2}{2} + q \phi  \sqrt{\frac{ (d - 3)}{ (d - 2)}\frac{S}{8 \pi}} ~R_2(\bar{g}_{ab}) + U(\phi )\right].
\end{equation} 
Notice that nothing in this derivation requires that the metric is a black hole.  The calculation is valid, locally on the transverse space, for any metric that is approximately the product of two-dimensional Rindler space and a transverse manifold, as in Eq.~\ref{eq:topologicalBH}.  The action above, with $U = 0$, is called the (free) Liouville action.

Differentiating the free Liouville action with respect to the metric and then setting it equal to the Rindler metric we find that the stress tensor of the $\phi$ field is \cite{solo} 
\begin{equation} 
T_{ab}  = \frac{1}{4}\left[ \partial_a \phi \partial_b \phi -  \bar g_{ab} (\nabla \phi)^2 \right] - q  \sqrt{\frac{ (d - 3)}{ (d - 2)}\frac{S}{8 \pi}}  \left[\partial_a \partial_b \phi -  \bar g_{ab} \nabla^2 \phi \right] .  
\label{eq:trace}
\end{equation} 
The trace vanishes when the flat space equation of motion $\nabla^2 \phi = 0$ is satisfied, verifying that this is a classical conformal field theory.  Importantly, this criterion is met in the near-horizon limit.  This is most easily seen by defining
\begin{equation} 
z = - \frac{1}{4\pi T_0} {\rm ln}\ (r - r_0) + z_0, 
\end{equation} 
where $r$ is the spatial Rindler coordinate in Eq.~\ref{eq:linearmetric}, and $z_0$ is a constant defined so that the horizon maps to $z = I/2$, with $I$ taken to infinity at the end of the calculation.  In this coordinate system, the blackening factor becomes $f(z) = f_0 e^\frac{2 z}{\beta_0}$.   Then the trace of Eq.~\ref{eq:trace} gives an equation of motion for $\phi$ \cite{solo}:
\beq
-\partial^2_t \phi + \partial_z^2 \phi = \frac{1}{4} q \Phi_0 f(z) R_2.
\eeq
The right-hand-side vanishes exponentially as $r \rightarrow r_0$, and in this limit the theory becomes conformal.  The non-vanishing trace can be viewed as the (classical) conformal anomaly of the flat space space Liouville theory in the background two dimensional metric, and it vanishes near the horizon because the metric becomes flat.

More importantly, the classical limit of the representation theory for the conformal spectrum of the free Liouville theory shows that all of the terms in $U(\phi )$ are relevant perturbations of the free Liouville Lagrangian.  The near-horizon limit is the analog of a UV fixed point.  We can zoom in on the horizon by rescaling the coordinate $ r - r_0$, which rescales the Rindler metric by a Weyl transformation.  The field $\phi$, which plays the role of an invariant entropy, has weight zero under this transformation.  Powers of $r - r_0$ higher than linear, as well as potential terms for $\phi$, scale away like relevant operators at an ultraviolet fixed point.  We emphasize that this is a purely classical argument, but we believe the analogy with the renormalization group makes it more transparent. Details of the conformal representation theory for the Liouville theory, with particular emphasis on the classical limit, can be found in~\cite{Seiberg} and references therein. 

We conclude that at the level of classical geometric symmetries, the generator $L_0$, which, according to Refs.~\cite{CHM,jv} becomes the modular Hamiltonian of a diamond in the quantum theory, is embedded in a Virasoro algebra.  Our conjecture, following Refs.~\cite{carlipreview,solo} is that the full Virasoro algebra is realized in the quantum theory, and that its representation satisfies the criteria for Cardy's theorem.\footnote{Knowledgeable readers will object that the Virasoro algebra (even its finite $sl(2,R)$ sub-algebra) has no finite dimensional unitary representations, contradicting the CEB for the diamond.  However, the considerations of~\cite{carlipreview,solo} and the present paper need only the subspace with $L_0$ in the vicinity of a finite value.  The true dynamics of quantum gravity will impose a cutoff on the Virasoro algebra of a diamond.  Nothing in this paper is sensitive to that cutoff. For example, our calculation of $\alpha = 1$ for a system with square root behavior of the log density of states is insensitive to cutoffs on the spectrum.}

Utilizing the $z-t$ coordinate system, which make the conformal symmetry manifest, the Poisson brackets of the dilaton
 \begin{equation} 
 \{\phi (z) , \partial_t \phi (z^{\prime}) \}_{PB} = \delta (z - z^{\prime}) ,  
 \end{equation} 
 allow one to write down the Poisson algebra of the charges $T_{++} = T_{00} + T_{0z}$ by identifying 
 \beq
 L_n = \frac{I}{2\pi} \int_{-I/2}^{I/2} dz e^{2\pi i n \frac{z}{I}} T_{++}.
 \eeq
 Here $I$ is the length of the finite interval on which the Liouville theory is defined.
%
%
The $L_n$ are, as a result, found to have a Virasoro algebra with central charge
 \begin{equation} 
 c = 6 q^2 \frac{d - 3}{d - 2} S. 
 \end{equation} 
Note that the free scalar stress tensor has vanishing classical central charge, but the canonical Poisson bracket relations
have induced a large central charge proportional to the square of the coefficient of the term linear in $\phi$. This classical central charge is much larger than the quantum correction to the central charge $\Delta c = 1$ of the free scalar, and the quantized $\phi$ field is not the correct description of the fluctuating degrees of freedom. 

To be able to use Cardy's formula and count the number of states $S_{\rm CFT} = 2\pi\sqrt{\frac{cL_0}{6}}$, one must have not only the central charge but also the value of $L_0$.  Applying Cardy's formula will allow us to determine $\langle K \rangle$ and $\langle \Delta K^2 \rangle$.  Since $L_0 = I^2 P^2/8\pi$, with $P = \partial_z \phi$, we need the field configuration.  The classical solutions of the Liouville model are right moving waves (the left moving waves don't contribute to $T_{++}$) and a zero mode solution which is a linear function of $z$.  The equilibrium configuration of the diamond is described by the zero mode.  
Solodukhin posits that the correct zero mode is
 \begin{equation}
  \phi = 2 \phi_0 \frac{z -  \beta_0}{ I-2 \beta_0} . 
  \end{equation} 
This obviously has the right value at the horizon, while the behavior at $z = 0$ is gotten by imposing that the boundary term in the EH action vanish there.\footnote{Since $z$ denotes the position of the horizon, which scales as $1/n$ in a replica calculation, one immediately sees that $S$ must scale inversely with the replica index $n$.  In a replica calculation, this would imply $\alpha = 1$.}    $L_0$ for the classical solution is then 
 \begin{equation}
  L_0 = \frac{\phi_0^2}{2\pi} ,
  \end{equation}
  for which
 Cardy's formula is 
 \begin{equation} 
 S_{CFT} = 2\pi\sqrt{\frac{cL_0}{6}} = S . 
 \label{eq:Cardy}
 \end{equation}
 Carlip derived the same result, including the case of rotating black holes, by purely $d$ dimensional methods.

It is a mistake to think that the quantized Liouville theory is the $1 + 1$ dimensional CFT to which the Cardy formula is being applied.  In that model the lowest value of $L_0$ is not the conformally invariant state, and Cardy's formula for the spectral density is not valid.  The density of states of the quantized Liouville theory is much too small to account for the entropy of the horizon. Moreover, the Liouville model is integrable, while the entanglement spectra of horizons are expected to be chaotic.  Instead, one should think of the results of Carlip and Solodukhin as a classical calculation of black hole entropy, analogous to that following from saddle point approximations to functional integrals over Euclidean metrics.  We expect, however, that a stochastic scalar field theory of $\phi$, with entropy given by Cardy's formula, and fluctuations satisfying $\alpha = 1$, could give an adequate effective description
of the full theory. We leave the precise formulation of such an effective description for future work.

It is useful to recall the calculation of Brown and Henneaux~\cite{brownhenneaux} which motivated the results of Carlip and Solodukhin.  Strominger~\cite{strominger} argued that the Brown-Henneaux calculation should be viewed as an avatar of the AdS/CFT correspondence.  The entropy of black holes in $AdS_3$ had been, in many cases, shown to be calculable in terms of $1 + 1$ CFT.  Earlier work of Brown and Henneaux showed that the asymptotic symmetry group of classical $AdS_3$ gravity contained a Virasoro algebra (in this case two sided) with a classical central charge.  Indeed, the boundary modes of $AdS_3$ gravity obey the equations of motion of the Liouville theory.  The classical central charge derived from $AdS_3$ gravity coincides with the actual central charge of the CFT describing AdS space according to the AdS/CFT correspondence. So the results of Carlip and Solodukhin should be viewed as a conjecture, analogous to the AdS/CFT conjecture, about the near-horizon quantum dynamics of black holes. The conjecture is that the modular Hamiltonian of a small black hole is the $L_0$ generator of a Virasoro algebra in a representation satisfying Cardy's theorem.  We have extended that conjecture to the spectrum of the modular Hamiltonian of any causal diamond with Rindler behavior near its horizon.

While we think this conjecture is extremely plausible, and have argued that Solodukhin's method generalizes to the near-horizon dynamics of {\it any} causal diamond with geometry $R_2 \times \Sigma_{d-2}$, for the purposes of calculating $\alpha$, we only need a more modest conjecture. 
The classical calculations of~\cite{carlipreview,solo} tell us that that the $L_0$ generator which preserves the diamond is embedded in a Virasoro algebra with a large central charge.  Since the modular Hamiltonian $K$ was also identified in Refs.~\cite{CHM,jv} as the generator of boosts that preserves the diamond,  
this suggests that $L_0$ should be identified with $K$.  We denote the spectrum of eigenvalues of $L_0$ (or equivalently $K$) as $E$, in which case the density of states, assuming that Cardy's theorem Eq.~\ref{eq:Cardy} is valid for the entropy of the quantum theory,
is given by $e^{B \sqrt{E}}$. We will assume a large constant $B$ such that we can carry out a saddle point approximation (this corresponds to large entropy).  
We then identify
\beq
\beta F_\beta =- \left( {\rm ln}\ \int_{E_1}^{E_2} dE\ e^{  B\sqrt{E} - \beta E}\right)
\eeq
in Eq.~\ref{eq:F},
from which one can calculate the modular fluctuations thermodynamically:
\begin{equation} 
\langle \Delta K^2\rangle  = \beta^2 \partial^2_{\beta}\left( {\rm ln}\ \int_{E_1}^{E_2} dE\ e^{  B\sqrt{E} - \beta E}\right)_{\beta = 1} . 
\end{equation} 
For large $B$, the integral is dominated by a stationary point on the contour of integration
\begin{equation} 
\frac{B}{\sqrt{E_*}} = 2 \beta .  
\end{equation}  
Taking just the saddle point contribution
we have
\begin{equation} 
\langle \Delta K^2 \rangle = \beta^2 \partial_{\beta}^2\left( \frac{B^2}{4\beta}\right)_{\beta = 1}  = \frac{B^2}{2\beta} . 
\end{equation}   
At the same time, the entropy is the microcanonical entropy at the saddle point,
\begin{equation} 
\langle K \rangle = B\sqrt{E_*} = \frac{B^2}{2\beta} , 
\end{equation} 
so that $\alpha = 1$.  For a $1 + 1$ dimensional CFT of course, the results $\alpha = 1$ follows from the fact that the entropy is linear in the temperature as a consequence of scale invariance and extensivity. This calculation shows that the result $\alpha = 1$ is more general than $1 + 1$ dimensional CFTs.  In particular, it shows that SYK models~\cite{sy,kitaev} (by which we mean Hamiltonians with fixed couplings belonging to one of the SYK ensembles) also have $\alpha = 1$ since they all have a density of states $\rho(E) \sim \sinh (B \sqrt{E})$\footnote{ SYK Hamiltonians with fixed couplings have a chaotic spectrum with no degeneracies. The average density of states emerges after coarse-graining this discrete spectrum over appropriate energy intervals. Similar remarks are valid for Cardy's formula, which applies to chaotic CFTs with no degeneracies.}.

We recall that for RT diamonds in AdS/CFT with an Einstein-Hilbert dual, the result $\alpha = 1$ was already obtained by multiple independent calculations~\cite{VZ2,perlmutter,Nakaguchi:2016zqi,deBoer:2018mzv}.  If one accepts the RS~II conjecture relating gravitational physics on a stabilized brane in AdS space to CFT dynamics, then our calculation in the RS~II model extends this result to causal diamonds in Minkowski space.  Our generalization of the conjecture of~\cite{carlipreview,solo} to arbitrary Rindler horizons implies that we should have $\alpha = 1$ for {\it local} fluctuations on any such horizon, including the cosmological horizon of dS space.  Below we will explain why this conjecture does not contradict the result $\alpha = d -2$, for large diamonds in AdS space, which follows from the thermal equilibrium fluctuations of large black holes.

Readers uncomfortable with the idea of attributing entropy to empty causal diamonds, should remember the fact that effective field theory {\it implies} the existence of a state independent\footnote{State independent, means {\it independent of the state in the Hilbert space constructed by completion of the space of states obtained by acting on the vacuum with a finite product of local operators}. Thermal states of the field theory do not lie in this Hilbert space.} entanglement entropy of any such diamond, whose leading term is proportional to $A(\Sigma_{d-2}) \Lambda^{d - 2}$, with $\Lambda$ a UV cutoff.    The map in Ref.~\cite{CHM} of this UV  divergence into a volume divergence on hyperbolic space shows that for an EFT flowing from a UV fixed point, the coefficient of this term is finite and calculable.  It depends on the UV fixed point, but in effective field theory the whole effect is a renormalization of $G_d$, if we accept the Covariant Entropy Bound, which attributes a classical gravitational entropy to the diamond boundary.
Thus we expect the universal entropy $A(\Sigma_{d - 2})/4 G_d$, for any small diamond in any finite model of quantum gravity\footnote{A version of this argument can be constructed using the AdS/CFT correspondence.  Polchinski~\cite{polch} and Susskind~\cite{suss} showed that for a CFT with an EH dual, one can construct states localized in a diamond much smaller than the AdS radius, by acting on the CFT with special combinations of local boundary operators $A_{i\ \diamond}$.   
The small diamond is described by a {\it code subspace}, which is maximally entangled with the much larger Hilbert space of the CFT.  Its density matrix, in generic states of the CFT, has high entropy, even though the diamond is empty. }. 

Our conjecture thus unifies effective field theory methods for calculating gravitational entropy~\cite{Sussug,carlipteitel,Callan:1994py,Cooperman:2013iqr,bdf} with the arguments of~\cite{carlipreview,solo}, and with the explicit calculations of~\cite{VZ2} for RT diamonds.  Like effective field theory replica methods, it is ultra-local 
on the $d - 2$ dimensional bifurcation surface.  Physically, one might think of all of these techniques as giving one the snapshot of horizon dynamics seen by a measuring device with Planck scale acceleration localized at a particular point on the bifurcation surface of the diamond boundary.  Ultra-locality on the bifurcation surface will be the key to understanding the discrepancy between the universal result $\alpha = 1$ and thermal entropy fluctuations for large AdS black holes.

To conclude this section we note that our conjecture is also suggested by the results of~\cite{Bredberg:2010ky}.  These authors showed that the viscosity to entropy density ratio for the hydrodynamic equations arising from Einstein's equations on any weakly curved diamond boundary, is universal, and equal to $\frac{1}{4\pi}$.  The Fluctuation-Dissipation theorem relates that ratio to the fluctuations of a hypothetical quantum operator, dual to the $T^{\tau i}$ components of the Brown-York stress tensor on a time-like near-horizon surface, in any model of quantum gravity in which Einstein's equations are approximately valid.  Our conjecture is the analogous statement for entropy fluctuations.  The hydrodynamics of entropy is {\it compressible} fluid dynamics, because of the thermodynamic relation for entropy density
\beq \sigma = (p + \rho)/T .\eeq We conjecture that the hydrodynamic equations equivalent to Einstein's equation are {\it incompressible} because of the fast scrambling~\cite{hp,ss} of information on certain null surfaces.  The dynamics that equilibrates entropy operates on a time scale $t_0 {\rm ln}\ S$ shorter than the diffusive time scale, $t_0$ times a power of $S$, which determines the flow of transverse momentum.  As a consequence, the entropy density in the Einstein/Navier-Stokes equations is constant.
This insight will be important in Section~\ref{sec:largediamonds}, where we seek to understand why the equilibrium fluctuations of large AdS black holes do not satisfy our universal rule.

 In order to relate the capacity of entanglement to a transport coefficient, and demonstrate the analog of the Fluctuation-Dissipation Theorem for entropy fluctuations, one would have to work out the analog of the Navier-Stokes equations that describes the fast scrambling of entropy on a small enough area of {\it any} stretched horizon.  This is the subject of future work.
 
 \section{Large Diamonds/Black Holes in AdS Space}
\label{sec:largediamonds}

We now turn to the exception that violates the rule, large diamonds or black holes in AdS space. 
For large black holes, the rules of the AdS/CFT correspondence say that they are dual to the thermal ensemble of a boundary CFT on a sphere.  The entropy and temperature computed by black hole thermodynamics are those of the density matrix  Eq.~\ref{eq:rho}, and as a consequence we can calculate the modular fluctuations via the thermodynamic formulae Eqs.~\ref{eq:K},~\ref{eq:DeltaKsq}, by taking a temperature derivative
$
\langle \Delta K^2 \rangle = T\frac{dS}{dT}.
$
Since $S \propto r_S^{d-2}$, it follows that we must find the temperature dependence of $r_S$.
For large mass, the Schwarzschild radius is given by 
\begin{equation} 
r_S^{d-1} = C_d M L^2, 
\end{equation} 
with $C_d$ determined from Eq.~\ref{eq:blackening}.
 The black hole temperature, given by $T = \frac{f'(r_S)}{4 \pi}$, is related to $L$ as
\begin{equation} 
T \propto (3-d)C_d \frac{M}{r_S(T)^{d-2}} + 2 \frac{r_S(T)}{L^2} = (5-d) \frac{r_S(T)}{L^2}. 
\end{equation}
The entropy is thus
$
S \propto T^{d-2},
$
and hence, 
\beq
\alpha = d - 2.
\eeq
As we discussed already in Sec.~\ref{sec:SecIII}, we can understand the failure of the $1 + 1$ dimensional CFT description of the horizon variables by noting that for large AdS black holes, the range of validity of the near-horizon approximation is small in comparison to the time required to span the horizon, since $r_S \gg L\pi / 2$, with the latter bounding the proper time of a generic time-like trajectory.

There is a feature of the dynamics that gives a clue to what is going on.  The spectrum of quasi-normal modes of a large AdS black hole includes sound modes, which indicate ballistic propagation of energy, momentum and information over the horizon~\cite{liu,rangamani}.  Note that there are no sound modes in the quasi-normal spectra of small diamonds, the dS horizon, or the boundary of an RT diamond. This is of course expected from the AdS/CFT correspondence: dynamics on the $d - 2$ sphere is described by a local field theory. Thus the temperature scaling of the entropy follows from scale invariance and extensivity of the entropy of a CFT on a boundary sphere $S^{d - 2}$ of radius $R$.  So, $\alpha = d - 2$ follows both from black hole thermodynamics, and from the same kind of scaling argument that we used for $1 + 1$ dimensional CFT, applied to $d-1$ dimensional CFT.  

A further piece of evidence is the fact that the sound modes on a large AdS black hole have a shortest wavelength, of order the AdS curvature radius $L$.  To see this, note that scale invariance of the boundary CFT, and the fact that the UV cutoff on sound modes is independent of the boundary volume, implies that the cutoff wavelength {\em on the boundary} is $\lambda_c \sim T^{-1} \sim \frac{L^2}{r_S} $.  The large black holes in AdS have a blackening factor 
\beq
f(r) = \frac{r^2}{L^2}-1 - \frac{r_S^{d-1}}{r^{d - 3} L^2}.
\eeq
%
The metric Eq.~\ref{eq:topologicalBH} with this blackening factor has a scaling symmetry~\cite{horhub}, which tells us that ratios of length scales on the black hole horizon are the same as those on the boundary.    So the cutoff length scale $l_c$ {\em on the horizon at radius $r_S$} satisfies
\begin{equation} 
l_c \sim \lambda_c \frac{r_S}{L} \sim L \ll r_S . 
\end{equation} 
On the other hand the spectrum of quasi-normal modes of shorter wavelength behave exactly like those on the horizon of a flat space black hole, because the geometries are identical on scales less than $L$.  The authors of~\cite{hp,ss} have argued that this behavior indicates fast scrambling of the information in a localized probe of some point on the horizon over a region of radius $L$, in a time $\sim L~ {\rm ln}(L/l_p)$.

The picture that one gets by combining these arguments is of a lattice of fast scrambling systems with $ \sim (L/l_p)^{d - 2} $ q-bits, and asymptotic densities of states like those of a $1 + 1$ dimensional CFT, coupled together in a manner local on the sphere.   
  The black hole entropy formula tells us that the modular fluctuations of large AdS diamonds are dominated by those of the $d-1$ dimensional ``sound modes".  This is to be expected.  The inequality $r_S \gg L$ tells us that the entropy of the full system
 scales like $(L/l_p)^{d - 2} (r_S / L)^{d - 2}$, which is much larger than the entropy of ``individual lattice points".

This picture is very natural in the context of tensor network regularizations of CFTs, which tile a spatial slice of AdS with a sequence of lattices.  The large coefficient of $T^{d-2}$ in the thermal entropy of CFTs with an EH dual implies that each of the nodes of the tensor network must be a Hilbert space of large dimension.  In this context, our conjecture states that the dynamics internal to each node are described by a model with the spectral density of a $1 + 1$ dimensional CFT.  Sound modes arise because of the couplings between different nodes.  The black hole entropy formula tells us that the entropy of the sound modes dominates the localized entropy at fixed temperature.   The authors of~\cite{Bredberg:2010ky} showed that the {\it incompressible} hydrodynamics of large black holes is identical to that of Minkowski diamonds in the near-horizon limit.  A localized perturbation of the large black hole, rapidly spreads its information over a node of the tensor network, on a time scale much shorter than the hydrodynamic scale, so the local entropy density appears constant.  It is only on length scales greater than the AdS radius that sound propagation becomes important.

To summarize, the quadratic thermal fluctuations of large black holes, which we claim are the same as the modular fluctuations of large empty diamonds in AdS space, are dominated by the behavior of the system on length scales much larger than the AdS radius.  They are proportional to the entropy, but with $\alpha = d -2$ rather than $1$.
On length scales below the AdS radius large AdS black hole horizons behave like other horizons studied in this paper.  It is important to realize that this reasoning does not apply to the boundary anchored RT diamonds studied in~\cite{VZ2}.  There, the proper time along time-like trajectories in the diamond are much larger than the AdS curvature radius; this allows ample time for information to propagate through the entire diamond.  Furthermore, there are no sound modes appearing at length scales above the AdS radius.

\section{Nested Causal Diamonds and the Random Walk}
\label{sec:nested}

\begin{figure}[btp]
\begin{center}
\includegraphics[scale=0.51]{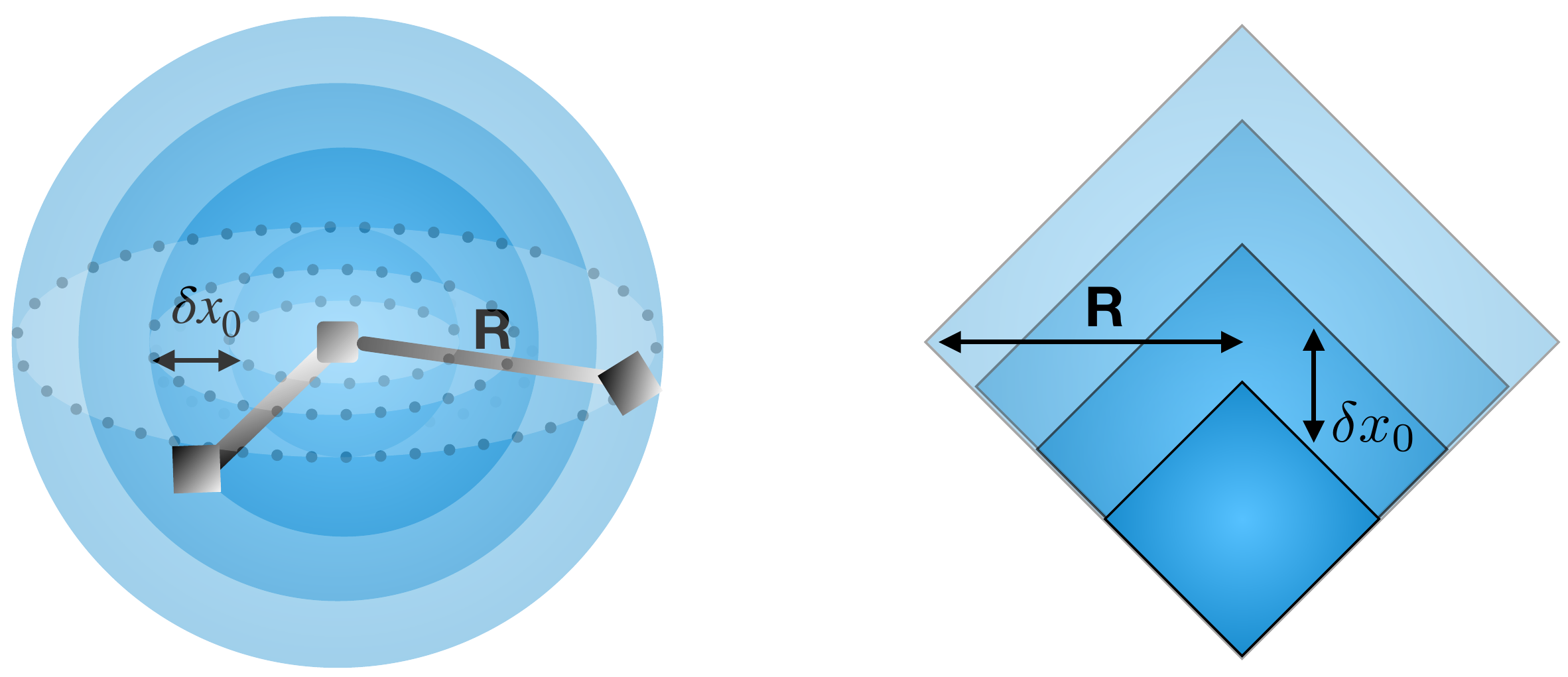}
\label{fig:causaldiamond}
\vspace{-0.5cm}
\caption{An interferometer of arm length $R$ traces out a series of nested causal diamonds, with each subsequent diamond separated by the scrambling length $\delta x_0$.  In the left panel, the nested causal diamonds are shown in position space, with each concentric sphere representing subsequent nested diamonds, while the right panel depicts the nested diamonds in the $x_i-x_0$ plane.}
\end{center}
\end{figure}

The results we have presented so far are mathematical formulae/conjectures applying to general causal diamonds.  In order to apply them to an actual interferometer experiment we must take into account the fact that the light beam in the apparatus passes through many causal diamonds.  More precisely, as depicted in Fig.~\ref{fig:causaldiamond}, we can cover the maximal causal diamond of the experiment by a nested sequence of diamonds (somewhat akin to Matryoshka dolls) corresponding to proper time intervals.   We must identify the interval $\delta x_0$ over which subsequent nested causal diamonds of size $x_0$ become statistically uncorrelated.  For simplicity we restrict ourselves to a spherically symmetric diamond in Minkowski space (relevant for an experiment), though the result can be generalized to RT and dS diamonds.   We calculate the entropy associated to a diamond at each subsequent step
\beq
S_{\rm ent} = \frac{\Omega_{d-2} (x_0 + \delta x_0)^{d-2}}{4 l_p^{d-2}} \simeq A_0 + \frac{\Omega_{d-2}(d-2)}{4}\frac{x_0^{d-2}}{l_p^{d-2}} \frac{\delta x_0}{x_0},
\eeq  
where the $d$-dimensional Planck length is defined here by $l_p^{d-2} =  G_d$, and $A_0$ is the entangling surface area when $\delta x_0 = 0$.  

We posit statistical independence when $\delta S_{\rm ent} \sim \sqrt{S_0}$, where $S_0$ in the entanglement entropy when $\delta x_0 = 0$.  This implies a scrambling length
\beq
\delta x_0 \simeq \frac{x_0}{(d-2)}\sqrt{\frac{4 l_{p}^{d-2}}{\Omega_{d-2} x_0^{d-2}}}.
\eeq
As the light beam in the interferometer travels to the mirror at a distance $R$ from the beam splitter, it traverses a sequence of statistically uncorrelated causal diamonds, the number of which is ${\cal N} = \frac{R}{\delta x_0}$.   The uncertainty in the length traversed, caused by fluctuations of the near-horizon degrees of freedom--the pixellons of Ref.~\cite{pixellon}--in {\it each} of these diamonds is of order $\delta x_0$, so the total length traversed should be thought of as a one dimensional random walk.  We thus have 
\beq
\delta R^2 \simeq \delta x_0^2 ~{\cal N} = \frac{R^2}{d-2}\frac{1}{\sqrt{S_0}}.
\eeq
This result is in agreement with the length fluctuations obtained in \cite{VZ2}, computed via a topological black hold foliation of a boundary-anchored diamond in AdS.  These length fluctuations can be thought of as the quantum width of the horizon of the causal diamond.
In $d=4$, this implies $\delta R^2 \sim l_p R$, an observably large fluctuation as proposed in Ref.~\cite{VZ1}.  In higher dimensions, the effect is suppressed by higher powers of $l_p$ and would be unobservable.

Note that an important aspect of our picture is that the transverse directions do {\em not} experience these length fluctuations, leaving the shape of the light cone intact.  This is important phenomenologically, because it implies that the images of stars do not become blurred over cosmological distances, a consequence of the entangling surface being minimal.  It is also a consequence, as we discussed above, of there being quasi-normal modes on Minkowski diamonds that are fast scramblers. We expect, as described in~\cite{VZ1}, that transverse fluctuations are given by an approximately Gaussian distribution with covariance $K(\Omega, \Omega^{\prime})$ that is rotationally invariant; such a distribution is described in terms of spherical harmonics in Ref.~\cite{VZ1}.  The calculations in this paper describe fluctuations only along the light cone. 

\section{Conclusions}

We have conjectured that the near-horizon dynamics of causal diamonds in flat space, boundary-anchored AdS and dS, are described in terms of a conformal field theory.   We focused on three pieces of evidence.  All three pieces of evidence imply $\langle \Delta K^2 \rangle = \langle K \rangle = \frac{A(\Sigma_{d-2})}{4 G_d}$, with the first two being explicit calculations and the third a heuristic argument.
First, we calculated $\langle \Delta K^2 \rangle = \frac{A(\Sigma_{d-2})}{4 G_d}$ for a causal diamond on a flat RS-II brane in the bulk of AdS.  
Second, we calculated the same from the density of states derived from the dimensional reduction of the Einstein-Hilbert action in flat space, following seminal work of Carlip and Solodukhin.
Third, we argued that the fluid-gravity correspondence combined with the fluctuation-dissipation theorem, as well as the nature of fast-scrambling horizons (in particular, the absence of sound modes for dS, Minkowski and boundary-anchored diamonds), supports $\alpha = 1$. 

We also showed that such an entropic (or holographic) picture of gravity implies that metric fluctuations in the causal development of a spherical ball ${\cal B}$ will {\em accumulate}, with each nested causal diamond (analogous to a Matryoshka doll, each with radius $\delta R \sim R/\sqrt{S}$ larger than the previous) uncorrelated with the previous causal diamond.  In $d=4$ dimensions, such trajectory-integrated fluctuations are observable in a future experiment, utilizing the advances in spacetime measurements achieved with gravitational wave observatories.

Taken together, these ideas imply that formal developments, often in the context of AdS or conformal field theories, need not be far removed from the realm of observation.  Possible concrete theoretical tools that could be utilized in the context of trajectory-integrated fluctuations are SYK models\footnote{We include SYK models because they also give rise to $\alpha = 1$. Strict adherence to the philosophy of~\cite{carlipreview,solo} would require us to restrict attention to CFTs.}, tensor networks and OTOCs that describe fast-scrambling systems.  We look forward to further development along these lines, including observational tests.

\newpage 
\begin{center}
{\bf Acknowledgments }\\
\end{center}
This paper is dedicated to the memory of M.~Virasoro, whose algebra proved pivotal.
We thank Patrick Draper for collaboration at the early stages of this work, Y. Chen, S. Gukov, V. Lee, D. Li for discussion, and S. Shenker, E. Verlinde, and especially C. Keeler and J. Parra-Martinez for comments on the manuscript.  TB thanks M. Rangamani, R. Myers and S. Carlip for email discussions of their work. The work of TB is partially supported by the Department of Energy under grant DOE SC0010008.  The work of KZ is supported by the Heising-Simons Foundation ``Observational Signatures of Quantum Gravity'' collaboration grant 2021-2817, by the DoE under contract DE-SC0011632, and by a Simons Investigator award.

\bibliography{QG}

\end{document}